\documentstyle[12pt]{article}
\def\PR #1 #2 #3 {Phys.~Rev.~{\bf #1}, #2 (#3)}
\def\PRL #1 #2 #3 {Phys.~Rev.~Lett.~{\bf #1}, #2 (#3)}
\def\PRD #1 #2 #3 {Phys.~Rev.~D~{\bf #1}, #2 (#3)}
\def\PLB #1 #2 #3 {Phys.~Lett.~{\bf B#1}, #2 (#3)}
\def\NPB #1 #2 #3 {Nucl.~Phys.~{\bf B#1}, #2 (#3)}
\def\RMP #1 #2 #3 {Rev.~Mod.~Phys.~{\bf #1}, #2 (#3)}
\def\ZPC #1 #2 #3 {Z.~Phys.~C~{\bf #1}, #2 (#3)}

\begin{document}

\rightline{DTP/95/40}
\rightline{ILL-(TH)-95-30}
\rightline{hep-ph/9505433}
\medskip
\rightline{May 1995}
\bigskip\bigskip

\begin{center} {\Large \bf Single-top-quark production via
$q\bar q \to t\bar b$} \\
\bigskip\bigskip\bigskip\bigskip
{\large\bf T.~Stelzer} \\ \medskip Department of Physics \\
University of Durham \\ Durham DH1 3LE \\ England \\
\bigskip\bigskip\bigskip
{\large\bf S.\ Willenbrock} \\ \medskip Department of Physics \\
University of Illinois \\ 1110 West Green Street \\  Urbana, IL\ \ 61801 \\
\bigskip
\end{center}
\bigskip\bigskip\bigskip

\begin{abstract}
We consider single-top-quark production via the weak process $q\bar q
\to t\bar b$ at hadron colliders.  This process may provide the best
measurement of the magnitude of the Cabbibo-Kobayashi-Maskawa matrix
element $V_{tb}$. We show that a signal can potentially be observed at
the Fermilab Tevatron with 3 $fb^{-1}$ of integrated luminosity.  In
contrast, the signal is masked at the CERN Large Hadron Collider by
top-quark pair production and single-top-quark production via
$W$-gluon fusion.
\end{abstract}

\addtolength{\baselineskip}{9pt}

\newpage

\section{Introduction}

\indent\indent The recent discovery of the top quark by the CDF and D0
Collaborations at the Fermilab Tevatron has stimulated interest in
top-quark physics \cite{TOP}.  The standard model of the strong and
electroweak interactions predicts that the top quark should behave,
both in its production and decay, as a heavy version of the five
lighter quarks.  However, the large mass of the top quark relative to
the five lighter quarks suggests that there may be something special
about its properties.  It is therefore crucial that we study the
properties of the top quark with every probe available to us.

In this paper we study a neglected probe of the top quark: the process
$q\bar q \to t\bar b$ via a virtual $s$-channel $W$ boson, as depicted
in Fig.~1(a) \cite{CP}.  This process is similar to the well-known
$W$-gluon fusion process \cite{DW,Y,EP,BE}, depicted in Fig.~1(b), in
that it produces a single top quark rather than a $t\bar t$ pair.
However, the process $q\bar q\to t\bar b$ probes the top quark with a
timelike $W$ boson, $q^2 > (m_t+m_b)^2$, while the $W$-gluon-fusion
process involves a spacelike $W$ boson, $q^2<0$.  These processes are
therefore complementary in that they probe the charged-current
interaction of the top quark in different regions of
$q^2$.\footnote{The decay of the top quark, $t\to Wb$, involves an
on-shell $W$ boson, $q^2=M_W^2$.}

Another important feature of $q\bar q \to t\bar b$ is that the
hadronic cross section can be reliably calculated.  The quark and
antiquark distribution functions are evaluated at moderate values of
$x$, where they are well-known.  The QCD correction to this process is
straightforward and can be carried out to at least ${\cal
O}(\alpha_s^2)$.  Furthermore, the initial quark-antiquark flux can be
constrained by measuring $q\bar q \to \bar\ell\nu$, which also
proceeds via a virtual $s$-channel $W$ boson.\footnote{Since the
neutrino longitudinal momentum cannot be reconstructed, the $q^2$ of
the $W$ boson cannot be determined.  The process $q\bar q \to
\bar\ell\nu$ therefore provides a constraint on the quark-antiquark
flux, not a direct measurement.}  The QCD correction to the initial
state is automatically taken into account via this procedure; only the
QCD correction to the final state must be explicitly calculated, and
this can be done reliably since it is free of collinear and infrared
singularities.\footnote{The QCD correction to the initial and final
states do not interfere until ${\cal O}(\alpha_s^2)$.}  In contrast,
the $W$-gluon-fusion process involves the gluon distribution function,
which is not well-known.

Since the most important issue to be resolved is the observability of
$q\bar q \to t\bar b$ above backgrounds, we concentrate on this
process in the context of the standard electroweak model.  In this
context, $q\bar q \to t\bar b$ may provide the best direct measurement
of the magnitude of the Cabbibo-Kobayashi-Maskawa (CKM) matrix element
$V_{tb}$.  If there are only three generations, unitarity of the CKM
matrix implies that $|V_{tb}|$ is very close to unity ($|V_{tb}| =
.9988-.9995$ \cite{PDB}).  If there is a fourth generation, $|V_{tb}|$
could be anywhere between (almost) zero and unity, depending on the
amount of mixing between the third and fourth generations.  A
measurement of $|V_{tb}|$ less than unity would therefore be a major
discovery.

\section{Signal and Backgrounds}

\indent\indent Including the top-quark decay, the process
$q\bar q \to t\bar b$ yields the final state $Wb\bar b$.  There are
several background processes which must be considered in order to
establish whether the signal can be observed.  The dominant
irreducible background, $q\bar q
\to Wb\bar b$, was considered in Ref.~\cite{CP}.  We reconsider this
background, as well as the other important irreducible and reducible
backgrounds.  The full set of backgrounds we consider is:
\begin{itemize}

\item $Wb\bar b$

\item $Wjj$ (where $j$ denotes a light-quark or gluon jet)

\item $WZ \to Wb\bar b$

\item $qg \to t\bar b j \to Wb\bar b j$ ($W$-gluon fusion)

\item $t\bar t \to W^+W^-b\bar b$

\end{itemize}
The raw cross sections for these processes at the Tevatron ($\sqrt s =
2$ TeV $p\bar p$ collider) and the CERN Large Hadron Collider (LHC,
$\sqrt s = 14$ TeV $pp$ collider) are given in the first column of
Table~1.  We use $m_t=175$ GeV, and all cross sections are calculated
at leading order.\footnote{The helicity amplitudes for the processes
in this paper are generated by MadGraph \cite{MADGRAPH}.
We use the MRS(A$'$) parton distribution functions \cite{MRSA},
evaluated at $\mu^2 = \hat s$, for all cross sections
except $W$-gluon fusion, where we use $\mu^2=M_W^2$.  We evaluate
$\alpha_s(\mu)$ at the same scale as the parton distribution
functions, with $\alpha_s(M_Z)=.117$ \cite{PDB}.}  The cross sections
include a factor of 2/9 for the branching ratio of $W \to \bar\ell\nu$
($\ell = e,\mu$). The cross sections listed are for $W^+$ production
only; the inclusion of $W^-$ doubles the signals and backgrounds at
the Tevatron, and nearly doubles them at the LHC. The $Wjj$ cross
section is infinite without cuts due to infrared and collinear
singularities.

\begin{table}[hbt]
\caption[fake]{Cross sections ($fb$) for $q\bar q \to t\bar b$ and a
variety of background processes at the Tevatron and the LHC. The cross
sections should be multiplied by a factor of two at the Tevatron, and
nearly a factor of two at the LHC, to include $q\bar q \to \bar tb$.
The $W$-gluon-fusion background is denoted by $t\bar b j$.  The first
column is the total cross section times a factor of 2/9 for the
branching ratio of the $W$ boson to $e,\mu$.  The second column adds
the cuts listed in Table~2 to simulate the acceptance of the detector.
The third column includes the rejection of events with an additional
identified jet or lepton.  The fourth column adds a cut on the
invariant mass of the $b\bar b$ pair. The final column includes the
efficiency for double $b$ tagging, taken to be $36\%$ per event (based
on $60\%$ efficiency per fiducial $b$ jet).  The fake rate for
light-quark and gluon jets is taken to be $1\%$ per jet.}
\bigskip
\begin{center}
\begin{tabular}{cccccc}
&&Tevatron&2 TeV $p\bar p$&&\\
\\
&Total$\times BR$&Acceptance&Rejection&$M_{b\bar b}>110$ GeV&$b$ tagging\\
\\
$t\bar b$  & 69   & 34    &    29 &   18 &  6.5
\medskip\\
$Wb\bar b$ & 3250 & 140   &   140 &   25 &  9.0 \\
$Wjj$      & --   & 14000 & 14000 & 3500 &  0.35\\
$WZ$       & 51   & 22    &    22 &  0.71&  0.26\\
$t\bar bj$ & 116  & 31    &   5.7 &  2.3 &  0.83\\
$t\bar t$  & 860  & 505   &   9.5 &  6.0 &  2.2 \\
\\
\\
&&LHC&14 TeV $pp$&&\\
\\
&Total$\times BR$&Acceptance&Rejection&$M_{b\bar b}>110$ GeV&$b$ tagging\\
\\
$t\bar b$ &  1100 &  290 & 190  & 130  & 47
\medskip\\
$Wb\bar b$& 34500 &  925 & 925  & 235  & 85 \\
$Wjj$     &    -- &340000&340000&140000& 14 \\
$WZ$      &   785 &  156 &  156 & 6.6  & 2.4\\
$t\bar bj$& 21000 & 4800 & 1100 & 615  & 220\\
$t\bar t$ & 90500 &40000 & 1200 & 810  & 290\\
\end{tabular}
\end{center}
\end{table}

\begin{table}[ht] \begin{center} \caption[fake]{Acceptance cuts used
to simulate the detector. The $p_{T\ell}$ threshold is greater for charged
leptons which are used as triggers (in parentheses).} \bigskip \begin{tabular}
{ll}
$|\eta_b|<2$ & $p_{Tb}>20$ GeV \\
$|\eta_\ell|<2.5$ & $p_{T\ell}>10$ GeV (20 GeV)\\
$|\eta_j|<4$ & $p_{Tj}>20$ GeV \\
$|\Delta R_{b\bar b}|>0.7 $ & $|\Delta R_{b\ell}|>0.7$ \\
${\not \!p}_{T}>20$ GeV & \\
\end{tabular} \end{center} \end{table}

We smear the jet energies with a Gaussian function of width
$\Delta E_j/E_j = 0.80/\sqrt {E_j} \oplus 0.05$ (added in quadrature) to
simulate the resolution of the hadron calorimeter.  We do not smear
the lepton energy, since this is a small effect compared with the
smearing of the jet energies.  We show in Fig.~2 the
transverse-momentum spectra of the $b$ jet, the $\bar b$ jet, and the
lepton for the signal process.  We list in Table~2 the cuts imposed on
the signal and backgrounds to simulate the acceptance of the
detector. The resulting cross sections are listed in the second column
of Table~1. The acceptance for the signal is $50\%$ at the Tevatron
and $25\%$ at the LHC.\footnote{The acceptance at the LHC will
actually be larger since the vertex detector extends to $|\eta_b|<2.5$
\cite{ATLAS,CMS}.}

In the third column of Table~1, the $W$-gluon fusion background
(denoted by $t\bar b j$) is reduced by rejecting events in which
a third jet is detected with $p_{Tj} > 20$ GeV and $|\eta_j|<4$.  This
provides a reduction factor of about 1/5 at both the Tevatron and the
LHC.  This also reduces the signal cross section, since the signal
will sometimes be accompanied by a third jet due to QCD radiation.  We
estimate\footnote{This estimate is derived from a calculation of
$t\bar b$ plus an additional jet, and an estimate of the QCD
correction to the inclusive cross section based on the correction to
$q\bar q\to \bar\ell\nu$.} this reduction to be about $15\%$ at the
Tevatron and about $35\%$ at the LHC, and reduce the signal
accordingly.\footnote{Since $W$-gluon fusion is a small background at
the Tevatron, it may be advantageous to increase the jet-rejection
threshold in order to preserve more of the signal.  With a threshold
of $p_{Tj}>30$ GeV, we estimate that almost none of the signal is
lost, while the small $W$-gluon fusion background is increased by only
a factor of 1.8.} The $Wb\bar b$, $Wjj$, and $WZ$ backgrounds would
also be reduced; but, to be conservative, we we have not taken
this into account.

The $t\bar t$ background is reduced by rejecting events with an
additional $W$ boson.  If the $W$ boson decays hadronically, the event
is rejected if either jet has $p_{Tj} > 20$ GeV and $|\eta_j|<4$.  If
the $W$ boson decays leptonically,\footnote{We treat the $\tau$
similar to $e,\mu$ for this $W$ decay. This may underestimate the
background for $W \to\bar\tau\nu$.} the event is rejected if
$p_{T\ell}>10$ GeV and $|\eta_\ell|<2.5$.  The overall reduction
factor from rejecting these events is 1/50 at the Tevatron and 1/30 at
the LHC.  Most of the remaining background events are from the
leptonic decay of the extra $W$ boson ($75\%$ at the Tevatron, $85\%$
at the LHC).\footnote{The threshold for rejecting leptons could
potentially be lowered below $p_{T\ell}>10$ GeV.  This is unnecessary
at the Tevatron, but is desirable at the LHC, where $t\bar t$ is the
largest background.}

In the fourth column of Table~1, we impose a cut on the $b\bar b$
invariant mass, $M_{b\bar b}>110$ GeV.  This essentially eliminates
the $WZ \to Wb\bar b$ background, and reduces the $Wb\bar b$ and $Wjj$
backgrounds.  The loss of signal is modest.\footnote{It may be
advantageous to relax this cut to preserve more of the signal.}  If
there is a Higgs boson of mass less than 120 GeV, it too will
contribute to the background via $WH \to Wb\bar b$ \cite{SMW,FR,MK},
and may require a slightly higher cut on $M_{b\bar b}$.

Due to the enormous background from $Wjj$ (where $j$ denotes a jet
from a light quark or gluon), it is necessary to tag one or both of
the $b$ quarks from the $q\bar q \to t\bar b \to Wb\bar b$ signal.
The most efficient means of $b$ tagging is with a silicon vertex
detector (SVX).  The CDF Collaboration has achieved a tagging
efficiency of roughly $40\%$ per ``fiducial'' $b$ jet ($p_{Tb} > 20$
GeV and within the coverage of the SVX), with a probability of less
than $1\%$ for a light-quark or gluon jet to fake a $b$ jet
\cite{TONY}. The new SVX for Run II, with three-dimensional vertex
information, should increase the efficiency to as high as $60\%$ while
maintaining or reducing the fake rate \cite{DAN}. Pixel detectors,
such as those planned for the ATLAS \cite{ATLAS} and CMS \cite{CMS}
detectors at the LHC, could further increase the efficiency.  We
assume a $60\%$ tagging efficiency\footnote{The soft-lepton tag also
contributes a small amount to the $b$-tagging efficiency.}  per
fiducial $b$ jet ($p_{Tb} > 20$ GeV, $|\eta_b| < 2$) with a $1\%$ fake
rate from fiducial light-quark and gluon jets. The best signal
significance is obtained by tagging both $b$ jets.\footnote{With
single $b$ tagging, the $W$-gluon-fusion background is prohibitively
large.  The $\bar b$ jet is often below the $p_T>20$ GeV threshold,
and the $b$ jet from the top decay plus the spectator jet (from the
radiation of the virtual $W$) fake the signal.}

The final column in Table~1 lists the cross sections including the
$b$-tagging efficiency.  With double $b$ tagging, the $Wjj$ background
is reduced to a negligible level at the Tevatron.

Another piece of information at our disposal is the invariant mass of
the $W$ boson plus $b$ jet, $M_{Wb}$, which equals $m_t$ for the
signal but has a continuous spectrum for the $Wb\bar b$ background.
Unfortunately, this information cannot be utilized directly.  One does
not know {\it a priori} which $b$ jet is from the top-quark decay.
Furthermore, the $W$-boson momentum must be reconstructed from the
lepton momentum plus the missing $p_T$, and this yields two solutions.
Thus each event has four possible assignments of momenta, leading to
four different values of $M_{Wb}$.

The most straightforward method to deal with this four-fold ambiguity
is to assign four values of $M_{Wb}$ to each event, and plot the
differential cross section as a function of this invariant mass
(weighting each entry by 1/4 of an event).  The resulting
invariant-mass distributions for the signal and the largest
backgrounds are shown in Fig.~3(a) at the Tevatron and Fig.~3(b) at
the LHC.  The signal is prominent at the Tevatron, but is masked by
the $t\bar t$ and $W$-gluon-fusion backgrounds at the LHC, which have
the same shape as the signal.  These backgrounds are enhanced at the
LHC relative to the $q\bar q\to t\bar b$ signal because they are
initiated by gluons.

In $p\bar p$ collisions, the $b$ jet from the process $q\bar q \to
t\bar b \to Wb\bar b$ tends to go in the direction of the proton beam
\cite{CP}.  This can be used to select the correct $b$ jet to
associate with the $W$ boson, with an efficiency of about $80\%$.
Furthermore, the correct neutrino solution tends to be the one with
the smaller absolute longitudinal momentum.  Making these assignments,
we show in Fig.~4 the resulting invariant-mass distribution at the
Tevatron.  The signal is significantly narrower than in Fig.~3(a), and
clearly stands out above the $Wb\bar b$ and $t\bar t$ backgrounds.

Selecting a bin of width 40 GeV centered on 175 GeV in Fig.~4, the
signal cross section is 4.4 $fb$ and the background 3.4 $fb$.
Including the charge-conjugate process, $q\bar q \to \bar tb$, the
total signal is 8.8 $fb$ on a background of 7.2 $fb$.  The number of
events acquired in 3 $fb^{-1}$ of integrated luminosity is enough for
the signal to correspond to more than a $5\sigma$ fluctuation of the
continuum background.  This integrated luminosity corresponds to about
1.5 years of running at the Tevatron with the Main Injector and the
Recycler (${\cal L}=2\times10^{32}/cm^2/s$) \cite{FOSTER}.

Assuming $|V_{tb}|$ is close to unity, 3 $fb^{-1}$ of integrated
luminosity yields a $10\%$ measurement of $|V_{tb}|$.  This
corresponds to a $20\%$ measurement of the partial width $\Gamma(t\to
Wb)$. A $5\%$ measurement of $|V_{tb}|$ requires 12 $fb^{-1}$, which
could be delivered in a little more than a year with a high-luminosity
Tevatron (${\cal L}=10^{33}/cm^2/s$) \cite{FOSTER}. This corresponds
to a $10\%$ measurement of $\Gamma(t\to Wb)$. The systematic error in
the measurement is small, as discussed in the Introduction.

If $|V_{tb}|$ is significantly less than unity, the rate for $q\bar q
\to t\bar b$ is suppressed, and this process is therefore more
difficult to detect.  If no signal is observed with 3 $fb^{-1}$ of
integrated luminosity, an upper bound of $|V_{tb}|<0.60$ ($95\%$
C.~L.) can be deduced.  The absence of a signal with 12 $fb^{-1}$
implies $|V_{tb}|<0.40$.

\section{Conclusions}

\indent\indent We have shown that the process $q\bar q \to t\bar b$ is
potentially observable at the Tevatron with 3~$fb^{-1}$ of integrated
luminosity. The signal is straightforward to isolate from backgrounds,
and requires tagging both $b$ jets. In contrast, this process is
masked at the LHC by backgrounds from $t\bar t$ and single-top-quark
production via $W$-gluon fusion.  We hope this work will motivate a
more detailed study of this process by the detector collaborations.

The process $q\bar q \to t\bar b$ at the Tevatron may be the best
direct measurement of $|V_{tb}|$.  The $W$-gluon fusion process also
provides a measurement of $|V_{tb}|$ \cite{Y}, and has the advantage
of a larger cross section.  However, it suffers from the uncertainty
in the gluon distribution function.  In contrast, the quark-antiquark
flux for $q\bar q \to t\bar b$ is well-known, and can be constrained
by measuring $q\bar q \to \bar\ell\nu$.  Both $q\bar q \to t\bar b$
and $W$-gluon fusion should be pursued, since they probe the top-quark
charged current in different regions of $q^2$.

\section*{Acknowledgements}

\indent\indent We are grateful for conversations with
C.~Hill, S.~Parke, and T.~Liss. T.~S. was supported in part by a
UK PPARC Post-Doctoral Fellowship. S.~W. was supported in part by
Department of Energy grant DE-FG02-91ER40677.

 \clearpage

\section*{Figure Captions}

\vrule height0pt \vspace{-22pt}

\bigskip

\indent Fig.~1 - Feynman diagrams for single-top-quark production via
(a) $q\bar q \to t\bar b$ via a virtual timelike $W$ boson, and
(b) $W$-gluon fusion via a virtual spacelike $W$ boson.

\indent Fig.~2 - Transverse-momentum spectra of the $b$, $\bar b$, and
lepton from $q\bar q \to t\bar b \to Wb\bar b$, followed by leptonic
decay of the $W$ boson, at the Tevatron.

\indent Fig.~3 - Differential cross section for $q\bar q \to t\bar b\to
Wb\bar b$ versus the invariant mass of the $W$ boson and $b$ jet at
(a) the Tevatron and (b) the LHC.  Also shown are the most important
backgrounds ($tbj$ denotes $W$-gluon fusion).  Each event contributes
four points to the distribution, each weighted by 1/4, due to the
ambiguity in the $b$ vs. $\bar b$ and the two-fold ambiguity in the
neutrino longitudinal momentum.  The signal is prominent at the
Tevatron, but is masked by the $t\bar t$ and $W$-gluon-fusion
backgrounds at the LHC.

\indent Fig.~4 - Same as Fig.~3(a), except the $b$ jet with the highest
rapidity with respect to the proton beam direction is selected as the $b$
from top decay, and the neutrino solution with the smaller absolute
longitudinal momentum is selected.  This narrows the distribution from
the signal in comparison with Fig.~3(a).

\vfill

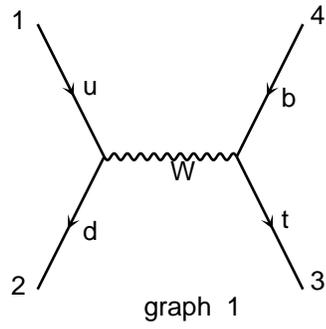
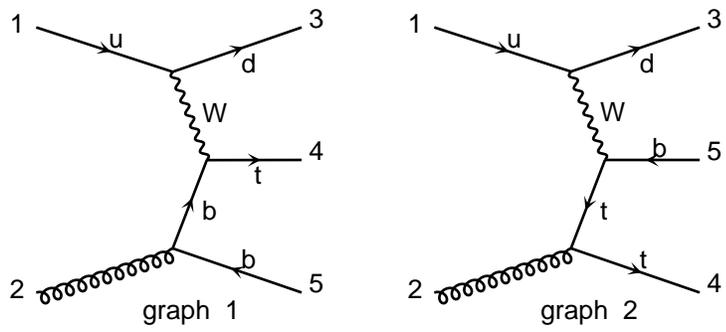
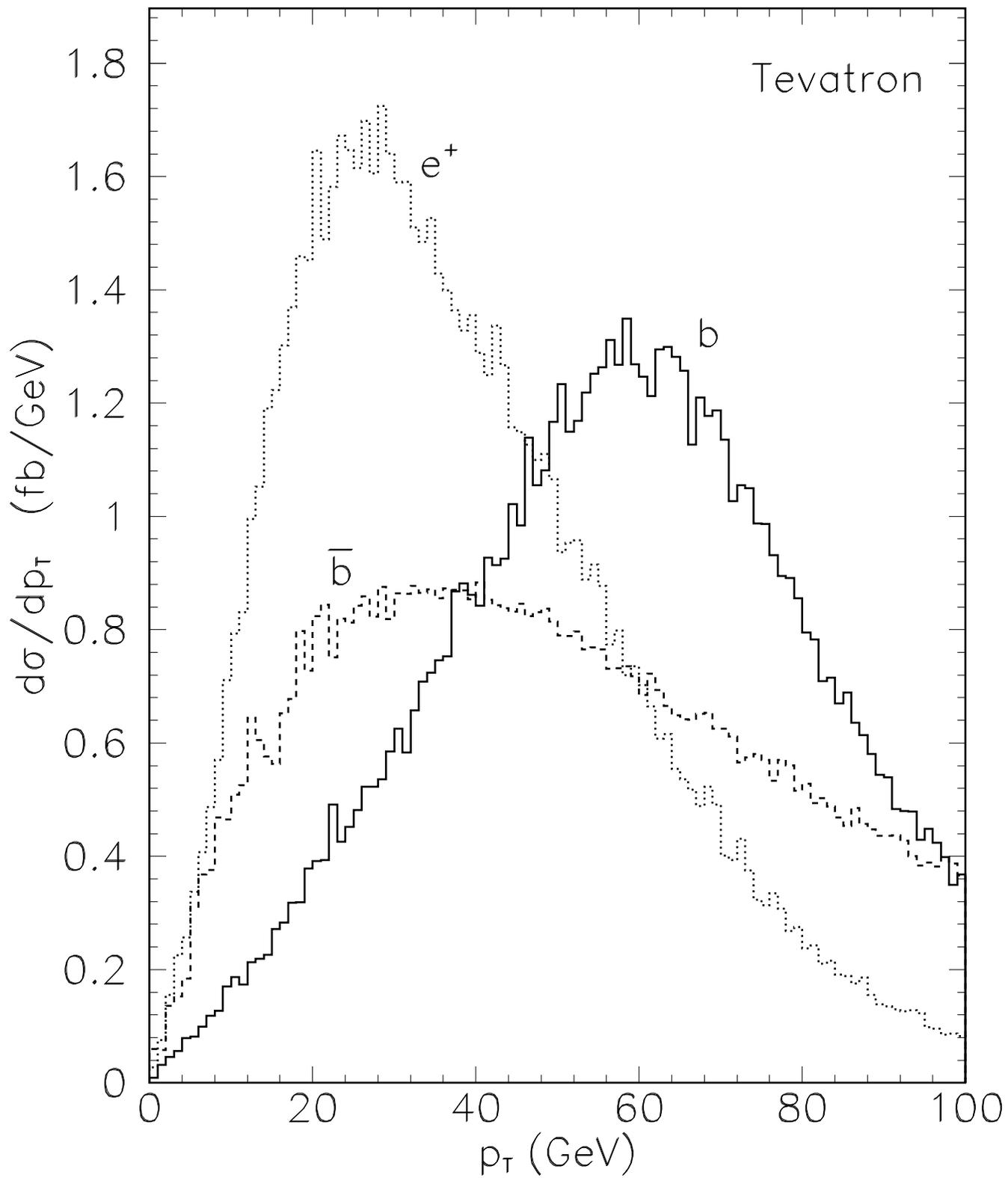
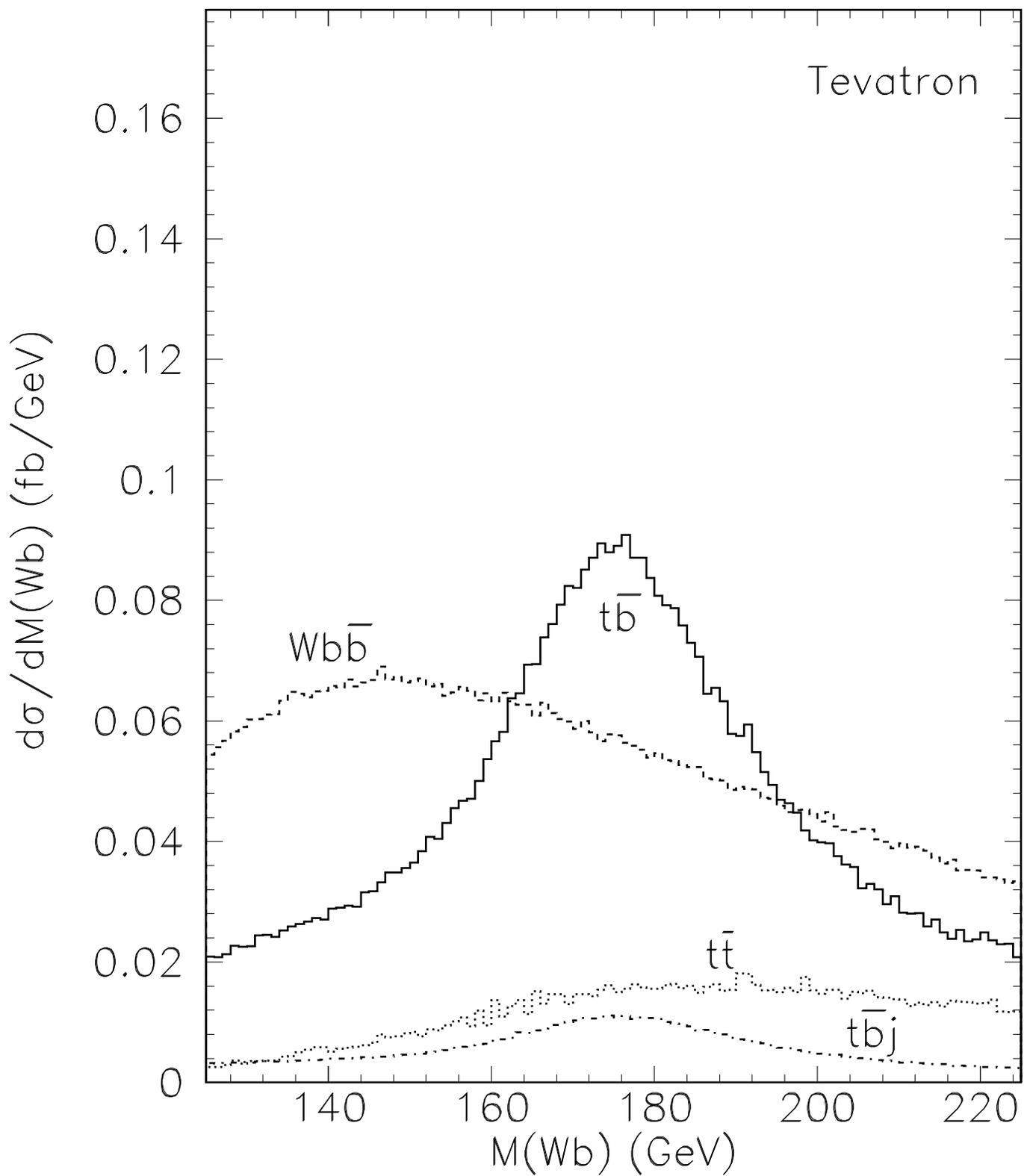
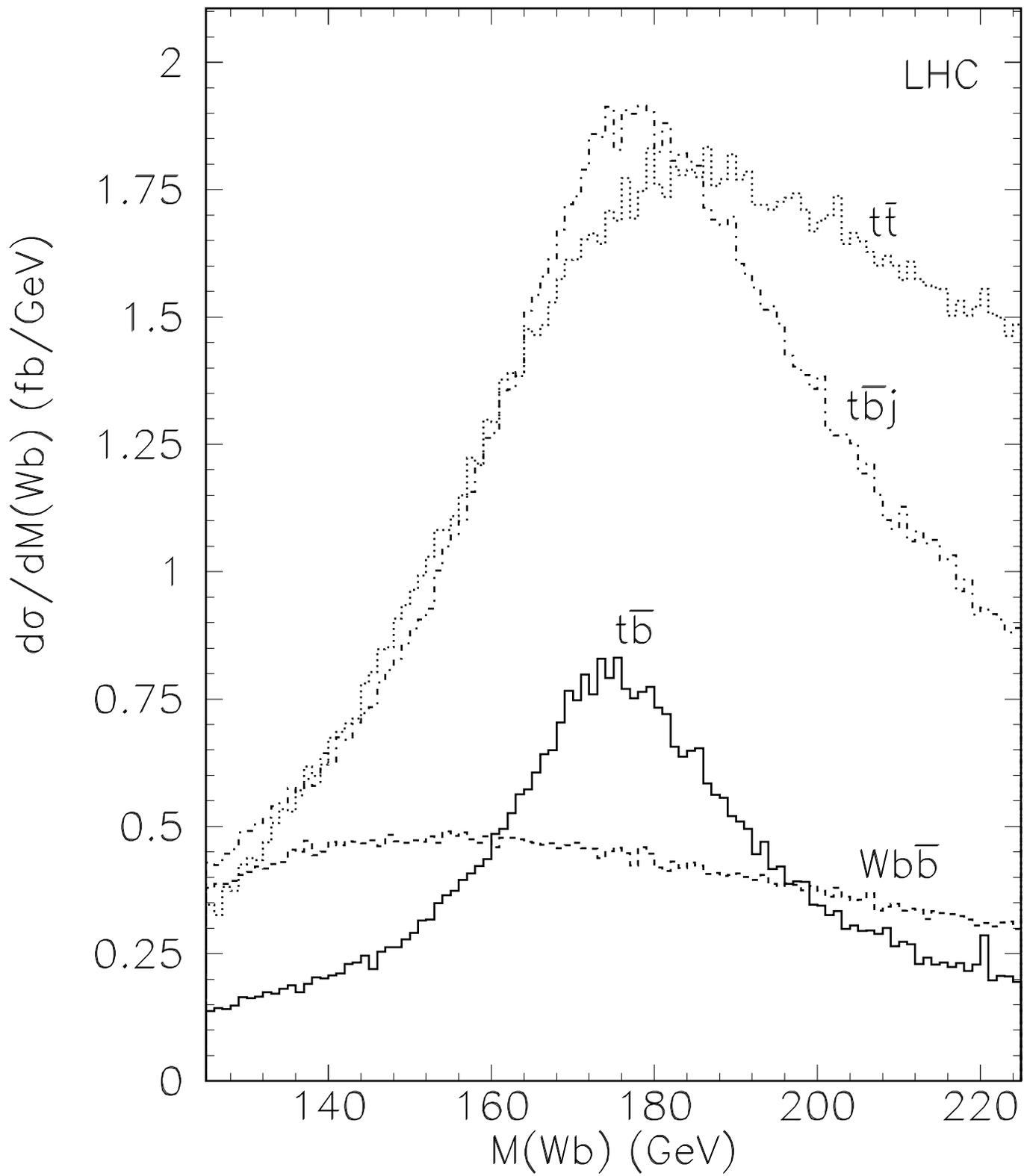
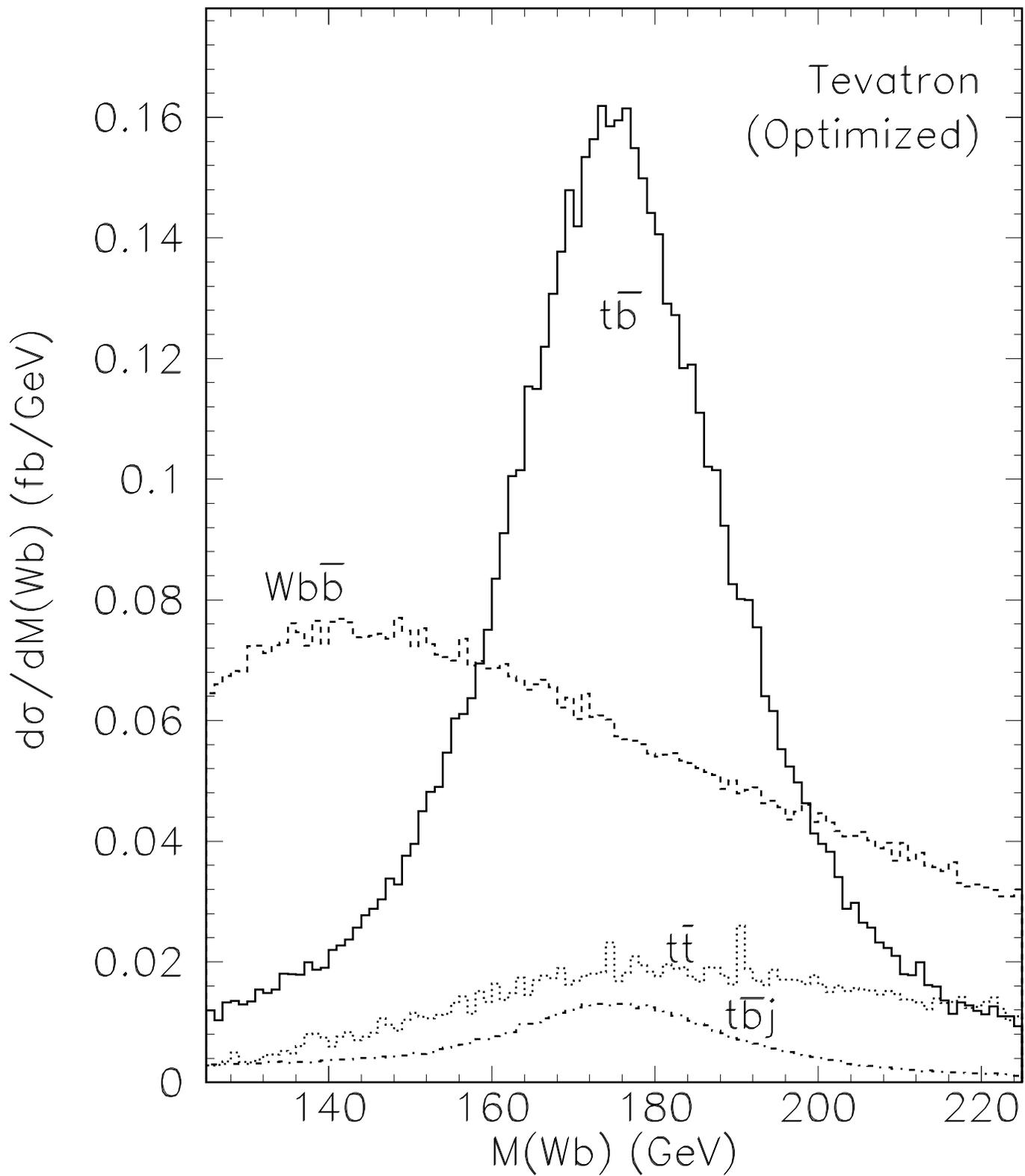
\end{document}